\documentclass{article}
\usepackage{graphics}
\usepackage{cite}

\def\be{\begin{equation}}
\def\ee{\end{equation}}
\def\bc{\begin{center}}
\def\ec{\end{center}}

\usepackage{amsmath}
\usepackage{theorem}

\usepackage{amsfonts,amssymb}
\usepackage{graphicx}

\begin{document}


\title{Fluctuations induce transitions in frustrated sparse networks}

\author{Adriano Barra\footnote{Dipartimento di Fisica, Sapienza Universit\`{a} di
Roma}}

\maketitle



\begin{abstract}
With the aim of describing a general benchmark for several complex
systems, we analyze, by means of statistical mechanics, a sparse
network with random competitive interactions among dichotomic
variables pasted on the nodes.
\newline
The model is described by an infinite series of order parameters
(the multi-overlaps) and has two tunable degrees of freedom: the
noise level and the connectivity (the averaged number of links).
\newline
We show that there are no multiple transition lines, one for every
order parameter, as a naive approach would suggest, but just one
corresponding to ergodicity breaking. We explain this scenario
within a novel and simple mathematical technique  via a driving
mechanism such that, as the first order parameter (the two replica
overlap) becomes different from zero due to a real second order
phase transition (with properly associated diverging rescaled
fluctuations), it enforces all the other multi-overlaps toward
positive values thanks to the strong correlations which develop
among themselves and the two replica overlap at the critical line.
\end{abstract}

\section{Introduction}

Among several different complex systems \cite{guido}\cite{SW} and
a large amount of tools for their investigation
\cite{barabasi}\cite{london}, statistical mechanics of disordered
systems earned an always increasing weight in the last two decades
\cite{amit}\cite{3}.
\newline
In this paper, the {\itshape complex networks} we analyze  by
statistical mechanics can be understood as follows: they are
networks because we allow the variables to live on the node of a
non trivial graph (a Poissonian Erdos-Renyi graph \cite{guido}),
the links among the nodes being the interacting fields they
exchange.
\newline
They are complex because, as opposite i.e. to the Ising model
\cite{abarra}
 (in which all the variables share the same coupling
constants) here the variables interact with equal probability via
a positive coupling or a negative one, giving rise to frustration
\cite{3} and forming what is often called, in the language of
statistical mechanics, a {\itshape diluted spin glass}
\cite{gt1}\cite{bray}, while, its zero temperature limit is known,
in the language of the theoretical computer science counterpart,
as a pairwise Random X-OR-SAT\cite{zecchina} (strictly speaking
random satisfability problems deal with p-spin models where
interactions happen in groups larger than couples \cite{mezard};
this is not a minor point as criticality in these systems is
related to the $p=2$ case, while for $p\geq3$ the phase transition
 is discontinuous\cite{barrap}\cite{bovier}, even though not first order in the sense of Ehrenfest \cite{landau} as there is no latent heat\cite{gardner}).
\newline
As these models are not Gaussian, they need not just a
(functional) order parameter (i.e. $q_2$)  as their fully
connected counterpart (i.e. the SK model  \cite{3}) but the whole
series of multi-overlaps (i.e. $q_2,q_4,...,q_{2n}$
\cite{gt1}\cite{bray}) and one may ask if there are several
transition lines for these multi-overlaps (one for every of them)
or they share the unique transition line at which ergodicity
breaks (the critical line for $q_2$). In a previous recent work
\cite{bdsf} we proved only mathematically, by bounds, the latter
scenario to be the correct one, but the physics behind was still
rather obscure and in particular no ideas concerning the nature of
this transition were presented.
\newline
In this paper  we show both mathematically (extending our previous
results) and physically (offering a picture for the nature of the
transition) a complete scenario as follows: At the boundaries of
the ergodic region the fluctuations of the first order parameter
(i.e. $q_2$) start diverging, accordingly to a well-defined second
order phase transition, while the fluctuations of all the others
do not (suggesting the validity of the several transition
alternative); however, due to the strong correlations that develop
at the critical point among all the order parameters, this growth
to a non zero value for $q_2$ drives all the others toward its
direction, acting as an 'ad hoc' field in the space of these
parameters. So the transition for the multi-overlaps surprisingly
is nor first order neither second order; it is a driven transition
via a self-generated coupling field which raises on the broken
ergodicity line.

\section{Equilibrium thermodynamics of the sparse frustrated network}

Consider $N$ nodes, indexed by Latin letters $i,j,$ etc., with an
Ising spin $\sigma_i = \pm 1$ attached to each of them. Let
$P_{\alpha N}$ be a Poisson random variable of mean $\alpha N$,
let $\{J_\nu\}$ be independent identically distributed copies of a
random variable $J$ with symmetric distribution. For the sake of
simplicity (but without loss of generality) we will assume $J=\pm
1$. We consider randomly chosen points, we therefore introduce
$\{i_\nu\},\{j_\nu\}$ as independent identically distributed
random variables, with uniform distribution over $1,\ldots,N$. The
Hamiltonian of the model (a suitable version of the Viana-Bray
\cite{bray} one) is the following symmetric random variable
\begin{equation}\label{ham} H_N(\sigma, \alpha; \mathcal{J})=
-\sum_{\nu=1}^{P_{\alpha N}} J_\nu \sigma_{i_\nu}\sigma_{j_\nu}\
,\ \alpha\in \mathbb{R}_+ \ .
\end{equation}
The non-negative parameter $\alpha$ is called {\sl connectivity}.
\newline
 The Gibbs measure $\omega$ and the partition function $Z_{N}(\beta)$ are defined by
$$
\omega(\varphi)=\frac{1}{Z}\sum_{\sigma}\exp(-\beta
H(\sigma))\varphi(\sigma), \ Z_{N}(\beta)=\sum_{\sigma}\exp(-\beta
H_{N}(\sigma))\ ,
$$
where $\varphi:\{-1,+1\}^N \to \mathcal{R}$ and $\beta$ is the
noise level in the network.
\newline
When dealing with more than one configuration, the product Gibbs
measure is denoted by $\Omega$, and various configuration taken
from each product space are called ``replicas''. $\mathbb{E}$ is
the expectation with respect to all the (quenched) variables, i.e.
all the random variables except the spins, collectively denoted by
$\mathcal{J}$ and we preserve the symbol $\langle . \rangle$ for
$\mathbb{E}\Omega(.)$. Sometimes we will deal with a perturbed
Boltzmann measure, whose perturbation is triggered by a tunable
parameter $t$ and we stress the dependence on such a perturbation
with a subscript $t$ on the averages $\langle . \rangle
\rightarrow \langle . \rangle_t$.
\newline
The (quenched) free energy density $f_{N}$ is defined by
$$
A_N(\beta,\alpha) = -\beta
f_{N}(\beta,\alpha)=\frac1N\mathbb{E}\ln Z_{N}(\beta,\alpha)\ .
$$
The whole physical behavior of the model is encoded by the even
multi-overlaps $q_{1\cdots 2n}$ \cite{bdsf}, which are functions
of several configurations $\sigma^{(1)},\sigma^{(2)},\ldots$ and
defined by
\begin{equation*}
q_{1\cdots 2n}=\frac1N
\sum_{i=1}^N\sigma_i^{(1)}\cdots\sigma_i^{(2n)}\ .
\end{equation*}
For the sake of simplicity, often we will denote by
$\theta=\theta(\beta)$ the expression $\tanh(\beta
J)=\tanh(\beta)$.

Looking for order parameter responses, in these networks, one
usually perturbs the system with a random field so to have \be
\tilde{H}_N(\sigma,h)= H_N(\sigma) + \sum_{i=1}^N h_i(t) \sigma_i,
\ee where the tilde stands for the perturbed Hamiltonian, $h_i$
are the random fields acting on the spins and $t \in [0,1]$ a
tuning of the amplitude of the perturbation, eventually sent to
zero afterwards (of course $h(0)=0$).
\newline
In our approach, due to the randomness of the coupling $J$ and the
gauge invariance of the model (the transformation $\sigma
\rightarrow \sigma\epsilon$, with $\epsilon \pm 1$ which leaves
the Hamiltonian unaffected being $\epsilon^2=1$) we can think at
the random perturbation as a term $h_i \sim
\sum_{\nu}^{P_{2\bar{\alpha}t}}\tilde{J}_{\nu}\sigma_{i_{\nu}}$
 then, by applying the gauge
$\sigma_{i_{\nu}}\rightarrow\sigma_{i_{\nu}}\sigma_{N+1}, \forall
i_{\nu}$, we can turn the perturbation into a cavity field,
mirroring an unperturbed system made by $N+1$ spins (whose
properties are the same of the $N$-spin system, for large $N$).
\newline
Notice that, thanks to the additivity property of the Poisson
variables,  we can also write, in distribution,
\begin{equation}\label{acca1}
H_{N+1}(\sigma; \alpha)\sim H_{N}(\sigma;\bar{\alpha})
+h_{\tau}\sigma_{1}, \ \ \ \bar{\alpha}=\alpha\frac{N}{N+1}\ ,
h_{\tau}=-\sum_{\nu=1}^{P_{2
\bar{\alpha}}}\tilde{J}_{\nu}\sigma_{k_{\nu}}\ .
\end{equation}
Let us define further a {\itshape cavity function}
$\Psi_{N,t}(\alpha,\beta)$ as the following quantity:
\begin{equation}
\Psi_{N,t}(\alpha,\beta) = \textbf{E}\ln \omega \big(
e^{\beta\sum_{\nu=1}^{P_{2\bar{\alpha}
t}}\tilde{J}_{\nu}\sigma_{i_{\nu}}} \big).
\end{equation}
Note that the cavity function takes into account the perturbation
applied to the original Hamiltonian; it plays a fundamental role
in the expansion of the free energy as it is immediately clear by
the next theorem \cite{barra}\cite{bdsf}:
\newline
\newline
\textbf{Theorem} {\itshape The following relation among free
energy, its connectivity increment and cavity function holds in
the $N\to \infty$ limit:}
\begin{equation}\label{FUS}
A_N(\alpha,\beta) + \alpha
\partial_{\alpha}A_N(\alpha,\beta) = \ln 2 +
\Psi_{N,t=1}(\alpha,\beta).
\end{equation}
The next two straightforward propositions  express explicitly the
two term by which the free energy can be decomposed thanks to eq.
(\ref{FUS}).
\begin{itemize}
\item  The incremental contribution to the free energy
by the connectivity is \cite{bdsf} \be\label{diana} \alpha
\partial_{\alpha}A(\alpha,\beta) = 2\alpha  \sum_1^{\infty}
\frac{1}{2n}\theta^{2n} (1 - \langle q_{2n}^2 \rangle). \ee
\item  The cavity function can be
represented by the integral of the series of all the fillable
multi-overlaps weighted by the powers of $\theta$ \cite{bdsf}:
 \be\label{malboro} \Psi_{N,t}(\beta,\alpha)=\int_0^t 2 \bar{\alpha}
\sum_{n=1}^{\infty}\frac{1}{2n} \theta^{2n}(\beta J)(1- \langle
q_{2n} \rangle_t')dt'. \ee
\end{itemize}
The next two propositions help us in understanding how to deal
with these two expressions:
\begin{itemize}
\item {\itshape Robustness} states that all the multi-overlaps which are
 ''filled" , i.e. they have each replica appearing an even number
 of times (like  $\langle q^2_{12} \rangle$, $\langle q^2_{1234} \rangle$, $\langle q_{12}q_{34}q_{1234} \rangle $) are not affected by the
 perturbation.
\newline
More sharply In the $N \to \infty$ limit, the average $\langle
\cdot \rangle_t$ of filled monomials is not affected by the
presence of the perturbation modulated by $t$, that is, for
instance,
$$
\int_{\bar{\alpha}_1}^{\bar{\alpha}_2}\langle
q_{12}q_{23}q_{13}\rangle_{t} d\bar{\alpha}=
\int_{\bar{\alpha}_1}^{\bar{\alpha}_2}\langle
q_{12}q_{23}q_{13}\rangle d\bar{\alpha}\ \ ,
$$
$\forall [\bar{\alpha}_1,\bar{\alpha}_2]$. We call this property
of filled monomials ''robustness" \cite{bds1}.

\item {\itshape Saturability} states that, once called "fillable" the other multi-overlap monomials,
in the $t \rightarrow 1, \ N \rightarrow \infty$ limits, fillable
monomials become filled  (i.e.
$\lim_{N\rightarrow\infty}\lim_{t\rightarrow 1}\langle q_{2}
\rangle_t = \langle q_{2}^{2} \rangle$,
$\lim_{N\rightarrow\infty}\lim_{t\rightarrow 1}\langle
q_{12}q_{34} \rangle_t = \langle q_{12}q_{34}q_{1234} \rangle $).
\newline
More sharply, let $q_{1\cdots 2n}$ be a fillable monomial of the
multi-overlaps, such that $q_{1\cdots 2n}Q_{1\cdots 2n}$ is
filled. Then
$$
\lim_{N \to \infty} \langle q_{1\cdots 2n}\rangle_{t=1} = \langle
q_{1\cdots 2n}Q_{1\cdots 2n} \rangle\ .
$$
We refer to this property as ''saturability" \cite{bds1}.
\end{itemize}
To obtain a stochastically stable and gauge invariant iterative
expression for the free energy, we have to expand the cavity
function via filled monomials: Neglecting orders higher than
$(2\alpha\theta^2)^{2}$ we  get
\begin{equation}
\Psi_{N,t}(\alpha,\beta)=\int_0^t dt'  \ 2
\alpha(\frac{\theta^2}{2}(1- \langle q_{12} \rangle_{t'}) +
\frac{\theta^4}{4}(1- \langle q_{1234} \rangle_{t'}) + ...).
\end{equation}
which can be filled by expanding its internal multi-overlap
monomials (i.e. \newline $\langle q_{12} \rangle_t =
2\alpha\theta^2 t \langle q_{12}^2 \rangle + O(t^2), \ \langle
q_{1234} \rangle_t = 2\alpha \theta^4 t \langle q_{1234}^2 \rangle
+ O(t^2)$) and than trivially integrated back thanks to
robustness.
\newline
We can now use Eq.(\ref{FUS}) to write down our free energy
expansion of the model. Presenting just the first orders, and
remembering that we call $\tau=2\alpha\theta^2$, we have
\begin{eqnarray}\label{freall} A(\alpha,\beta) = \ln 2 &+&
(\frac{1}{2\alpha})^0\Big(
\frac{\tau}{2}-\frac{\tau}{4}(1-\tau\theta^0)\langle q_{12}^2
\rangle  + \frac{\tau^3}{3}\langle q_{12}q_{23}q_{13}\rangle +
...\Big) \\ \nonumber &+&
(\frac{1}{2\alpha})^2\Big(\frac{\tau}{4}-\frac{\tau}{8}(1-\tau\theta^2)\langle
q_{1234}^2 \rangle +\frac{3\tau^3}{4} \langle q_{1234}q_{12}q_{34}
\rangle + ...\Big) + ...\end{eqnarray} Note that in the high
connectivity limit \cite{gt1} the expression (\ref{freall})
approaches the well known expression for the free energy of the SK
model \cite{barra}\cite{3}.

\section{Order parameter fluctuations and uniqueness of critical line}

The multi-overlaps among any $2n$ configurations is typically
small in the ergodic region defined by $2\alpha\tanh^2(\beta)=1$
and their fluctuation can be studied on the $\sqrt{N}$ scale by
defining  \be \eta_{2n}= \sqrt{N}q_{2n} =
\frac{1}{\sqrt{N}}\sum_i^N \sigma_i^1...\sigma_i^{2n}. \ee Then it
is possible to show that these rescaled multi-overlaps behave, in
this region, like independent centered Gaussian variables, in the
infinite volume limit, and the following theorem holds \cite{gt1}:
\newline
\newline
\textbf{Theorem} {\itshape In the annealed region
$2\alpha\tanh^2(\beta)<1$ the variables $\eta_{2n}$ converge to
centered Gaussian process with covariances
\begin{eqnarray}\label{tonino}
\langle \eta_{a_1,...,a_{2n}} \rangle &=&
\frac{1}{(1-2\alpha\mathbb{E}\tanh^{2n}(\beta J))} \\
\langle \eta_{a_1,...,a_{2n}}\eta_{b_1,...,b_{2n}} \rangle &=& 0 \
\ if \ \ \exists i : a_i \neq b_i
\end{eqnarray}
and, when the boundary of the annealed region is approached,
{\itshape only} the variance of $\eta_2$ diverges.}
This theorem for the  fluctuations of $q_2$ and for finding its
critical line is straightforward within our method so we sketch
the proof:
\newline {\itshape Sketched proof} At first we
 expand the $2$-replica overlap \be\label{prima} \langle q_{12}
\rangle_t = 2\alpha\theta^2 \langle q_{12}^2 \rangle
-4\alpha^2\theta^4 \langle q_{12}q_{23} \rangle_t + O(q^3). \ee
Then, by simple polynomial integrations, we can evaluate the
overlap expansion in terms of filled monomials. \be \langle q_{12}
\rangle_t = 2\alpha\theta^2\langle q_{12}^2 \rangle t -4
\alpha^2\theta^4 \int_0^t dt'\int_0^{t'} dt'' \langle
q_{12}q_{23}q_{13} \rangle + O(q^6). \ee Now, by applying
''saturability", we get $\langle q_{12} \rangle_t = \langle
q_{12}^2 \rangle$, consequently, forgetting $O(q^{4})$ terms and
multiplying by $N$, we have \be\label{PC1} \langle \eta_2^2
\rangle =
\frac{2(2\alpha\theta^2)^2}{(1-(2\alpha\theta^2))}\langle
\eta_{12}\eta_{23}\eta_{13} \rangle. \ee  We see that at the
r.h.s. the overlap order is $3$ while at the l.h.s. is $2$: By a
Central Limit Theorem argument we see that the only diverging
point, for the rescaled overlap fluctuations is
$2\alpha\theta^2=1$, where the r.h.s. denominator explodes $\Box$.
\newline
To try and show our physical picture, let us start by the
following theorem:
\newline
\newline
\textbf{Theorem} {\itshape Given two integer numbers $c,d$  such
that $cd=2n$ and $m\in N$ the following families of bounds hold
generically and also at finite $N$:} \be\label{bounds} \langle
q_{2n}^m \rangle \geq \langle q_{1..c}^{m}
q_{c+1..2c}^{m}...q_{c(d-1)+1..2n}^{m} \rangle \geq \langle
q_{1..c}^m \rangle^d \ee {\itshape Sketched
 proof} Always using $q_2$ and $q_4$ as examples, we prove the
theorem for $c=d=2$ and $m=1$. Its generalization is
straightforward.
\newline
Exploiting the factorization of the Boltzmann state at fixed $J$
one has \be\nonumber \langle q_{1234}\rangle
  =  \mathbb{E}\frac{1}{N}\sum_i \omega^4(\sigma_i)
\geq \mathbb{E}(\frac{1}{N}\sum_i \omega^2(\sigma_i))^2
 = \mathbb{E}\omega^2(q_{12})
\geq (\mathbb{E}\omega(q_{12}))^2
 = \langle q_{12} \rangle ^2,
\ee where we have used  $\mathbb{E}[a^2] \geq \mathbb{E}^2[a]$ for
any real-valued random variable, first for $a=\omega^4(\sigma_i)$
and with the expectation taken over the uniform distribution on
$i={1,...,N}$ and then for $a=\omega(q_{12})$ with the expectation
over $P(J)$.$\Box$
\newline
The conclusion is that it is not possible to have several spin
glass transitions in any model: as soon as $\langle q_{12}
\rangle$ becomes nonzero, also $ \langle q_{1234} \rangle$ must
be, and so on.
\newline
The mechanism we provide is again ultimately based on
saturability. In fact at the critical point the fillable
multi-overlap $\langle q_{12}q_{34} \rangle$, applying
saturability, gets \be \lim_{N\rightarrow
\infty}\lim_{t\rightarrow 1}\langle q_{12}q_{34} \rangle_t =
\langle q_{12}q_{34}q_{1234} \rangle, \ee which couples the first
multi-overlap $q_2$ and the second multi-overlap $q_4$ together,
generating the correlation which drives the transition for
$\langle q_{1234}\rangle$. Saturability can be applied as we are
at the boundary of the ergodicity breaking (the last point in
which it still holds due to  a real second order phase transition
of $q_2$).

So remembering once more that we are taking just the first two
multi-overlaps but the scheme applies to all them and, for the
sake of the clearness consequently forgetting all the higher order
not necessary terms, we can write the free energy, that we call
$f(q_2,q_4)$ stressing the dependence by the two multi-overlaps as
\be f(q_2,q_4) = (\theta-(\frac{1}{2\alpha})^{\frac{1}{2}})q_2^2 +
(\theta-(\frac{1}{2\alpha})^{\frac{1}{4}})q_4^2 -
\frac{3\tau^3}{4} q_2^2 q_4 \ee and we want to know how the minima
of $f(q_2,q_4)$ evolve with $\theta$ (at fixed $\alpha$, or
viceversa). If a bifurcation analysis of the saddle point
equations from the origin is performed, one would find two
transition lines, $\theta_{q_2}=(1/2\alpha)^{1/2}$ and
$\theta_{q_4}=(1/2\alpha)^{1/4}$. However, when looking  at the
actual minima it is possible to see just the first transition.
After that the two minima are away from the origin and so the
second "potential transition line" at
$\theta_{q_4}=(\frac{1}{2\alpha})^{\frac{1}{4}}$ never appears:
when approaching this line the system is already in a completely
different part of its phase space. We stress that above
$2\alpha\theta^2=1$, where the quadratic expansion of $f(q_2,q_4)$
around the origin determines the Gaussian fluctuations, $q_2$ and
$q_4$ are uncorrelated, than, below this line, the third-order
term produces an interaction ($q_{12}q_{34}q_{1234}$) and so, as
soon as  $q_2$ becomes non zero, it also drives $q_4$ to a non
zero value. It is also straightforward to check that near
$2\alpha\theta^2$ the minima scale as $q_2 \sim
(2\alpha\theta^2-1)^{1/2}, q_4 \sim (2\alpha\theta^2) \sim q_2^2$
accordingly with the proved scaling for random spins at
criticality \cite{bdsf}.

\section{Summary}

In this paper we analyzed the genesis of the phase transition in
frustrated sparse networks, by matching a rigorous approach
(essentially based on modern cavity interpolation \cite{barra})
with a theoretical picture (essentially provided via replica trick
\cite{bray}). Overall a clear scenario for the transition in these
systems has been achieved: at the onset of ergodicity breaking the
first order parameter (i.e. $q_2$) undergoes a second-order phase
transition; due to the correlations among this parameter and all
the others (i.e. $q_4$), it drives the latter to a positive value
too. The positivity of the values assumed by these parameters
(another prescription of Parisi theory \cite{3}) is a
straightforward application of the saturability property on
themselves. This has interesting consequences, ranging from
disordered statistical mechanics to computer science as well as
random matrix theory. On the same line, we stress that in recent
years, even on the last subject\cite{bray2}, an increasing
formalization (avoiding replicas), from Girko's framework
\cite{girko}, has been achieved \cite{isaac}.

\bigskip

{\bf ACKNOWLEDGEMENTS.} Peter Sollich is warmly acknowledged for
illuminating discussions.


\begin{thebibliography}{99}

\bibitem{amit} D.J. Amit {\em Modeling brain functions}, Cambridge University Press, (1992).

\bibitem{barra} A. Barra,
  \emph{Irreducible free energy expansion for mean field spin glass model},
  J. Stat. Phys. \textbf{123} (2006).

\bibitem{abarra} A. Barra, {\em The mean field Ising model trough interpolating
techniques}, J. Stat. Phys. \textbf{132} (2008).

\bibitem{barrap} A. Barra, {\em Notes on the ferromagnetic P-spin and
REM}, Math. Meth. in Appl. Sc. $10.1002/mma1065$, Wiley, (2008).

\bibitem{bds1} A. Barra, L. De Sanctis
  \emph{Stability properties and probability distributions of multi-overlaps in dilute spin
  glasses}, J. Stat. Mech. P08025 (2007).

\bibitem{bdsf} A. Barra, L. De Sanctis, V. Folli
  \emph{Critical behavior of random spin systems}, J. Phys. A: Math. Theor. \textbf{41} No 21 215005,
  (2007).

  \bibitem{bovier} A. Bovier, I. Kurkova, M. Loewe, {\em Fluctuations of the free energy in the
REM and the p-spin SK model}, Ann. Probab. \textbf{30}, (2002).

\bibitem{guido} G. Caldarelli, A. Vespignani, {\em Large scale
structure and dynamics of complex networks}, World Scientific
Publishing, (2007).

\bibitem{gardner} E. Gardner, {\em Spin glasses with p-spin interactions}, Nucl. Phys.
B, \textbf{257}, 747 (1985).

\bibitem{girko} V.L.Girko, {\em Spectral Theory of Random
Matrices}, Nauka, Moscow, (1988).

\bibitem{gt1} F. Guerra, F.L. Toninelli, \emph{The high temperature region
    of the Viana-Bray diluted spin glass model},
  J. Stat. Phys.Ê \textbf{115}  (2004).

\bibitem{landau} L.D. Landau, E.M. Lifshitz, {\em Statistical Physics, Part \textbf{1}}, Vol. \textbf{5}, Course of Theoretical Physics, Pergamon, 3rd Ed. (1994).

\bibitem{mezard} M. Mezard, T. Mora, R. Zecchina, {\em Clustering of solutions in the random satisfiability
problem}, Phys. Rev. Lett. \textbf{94}, 197205, (2005).

\bibitem{3}M. Mezard, G. Parisi, M.A. Virasoro, {\em Spin Glass Theory and Beyond}, World scientific publishing,
(1987).

\bibitem{zecchina} M. Mezard, F. Ricci-Tersenghi, R. Zecchina,
{\em Alternative solutions to diluted p-spin models and XORSAT
problems}, J. Stat. Phys. \textbf{111}, (2003).


\bibitem{barabasi} M. Newman, D. Watts, A.L. Barabasi {\em
The Structure and Dynamics of Networks}, Princeton University
Press, (2006).

\bibitem{bray2} G.J. Rodgers, A. Bray, {\em Density of states of a sparse random matrix}, Phys. Rev. B \textbf{37}, 3557,
(1988).

\bibitem{isaac} T. Rogers, K. Takeda, I. Perez-Castillo, R. Kuhn, {\em Cavity Approach to the Spectral Density of Sparse Symmetric Random
Matrices}, Phys. Rev. E. \textbf{78}, 031116, (2008).

\bibitem{london} P. Sollich, A.C.C. Coolen (eds): {\em Proceedings
in Disordered and Complex Systems}, King's College London, UK,
 (2000).

\bibitem{bray} L.Viana, A.Bray, {\em Phase diagrams for diluted spin
glasses}, J. Phys. C, \textbf{18}, (1985).

\bibitem{SW} D.J. Watts , S.H. Strogatz {\em Collective dynamics of 'small-world' networks}, Nature, \textbf{393},  (1998).


\end{thebibliography}
\end{document}